\documentclass[aps,amsmath,amssymb,prl,fleqn]{revtex4}
\usepackage{graphicx}
\usepackage{latexsym}
\usepackage{amssymb}
\usepackage{amsmath}
\newcommand{\BE}{\begin{equation}}
\newcommand{\EE}{\end{equation}}
\newcommand{\BA}{\begin{eqnarray}}
\newcommand{\EA}{\end{eqnarray}}
\newcommand{\ND}{\noindent}
\begin{document}

\title{Nonspreading wave packets in a general potential V(x,t) in one dimension}
\author{Chyi-Lung Lin}  \author{Meng-Jie Huang} \author{Te-Chih Hsiung}\affiliation{Department of Physics, Soochow University,\\
Taipei, Taiwan, 111, R.O.C.}
\date{\today}

\begin{abstract}
We discuss nonspreading wave packets in one dimensional
Schr\"{o}dinger equation. We derive general rules for constructing
nonspreading wave packets from a general potential
$\textmd{V}(x,t)$. The essential ingredients of a nonspreading
wave packet, the shape function $f(x)$, the motion $d(t)$, the
phase function $\phi(x,t)$ are derived. Since the form of the
shape of a nonspreading wave packet does not change in time, the
shape equation should be time independent. We show that the shape
function $f(x)$ is the eigenfunction of the time independent
Schr\"{o}dinger equation with an effective potential
$V_{\textmd{eff}}$ and an energy $E_{\textmd{eff}}$. We derive
nonspreading wave packets found by Schr\"{o}dinger, Senitzky, and
Berry and Balazs as examples. We show that most stationary
potentials can only support stationary nonspreading wave packets.
We show how to construct moving nonspreading wave packets from
time dependent potentials, which drive nonspreading wave packets
into an arbitrary motion.


\end{abstract}


\maketitle
\newpage
\section{Introduction}
Nonspreading wave packets are those packets whose form of the
probability density function $|\Psi(x,t)|^2$ under time evolution
do not change. The shapes of energy eigenstates of a time
independent potential are stationary and are therefore
nonspreading in the trivial case. The first interesting
nonspreading wave packets was found by Schr\"{o}dinger \cite
{Schrodinger}. This packet is normalizabe, and is the shifted
ground state of a harmonic oscillator. The motion of the packet
behaves just like a classical particle shifted from its
equilibrium position, executing the simple harmonic motion. After
that, Senitzky extended the result showing that shifted higher
energy eigenstates are also nonspreading \cite{Senitzky}. Senitzky
also showed that the expectation value of energy of a nonspreading
wave packet is $E_n + E_{cl}$, where $E_n$ is the quantum
mechanical energy of a wave packet when it is stationary, and
$E_{cl}$ the classical energy of the particle. Thus, though the
motion of a classical particle and a nonspreading wave packet is
the same, their energy contents are different. We may view $E_n$
the structure energy of the stationary wave packet besides the
classical energy.

The extension of nonspreading packets from shifted ground state to
shifted higher energy eigenstates were also found by Roy and Singh
\cite{Roy},  Yan  \cite{Yan}, and Mentrup \textit{et al}.
\cite{Mentrup}, etc.

Much later than Schr\"{o}dinger, Berry and Balazs found another
interesting nonspreading Airy packet $\textmd{Ai}[x]$ in free
space \cite{Berry}. Airy packets $\textmd{Ai}[x]$ though bounded
are not square integrable. An Airy packet does not move in uniform
velocity in free space, instead, it accelerates. Thus, in contrast
to Schr\"{o}dinger packets, the motion of an Airy wave packet is
different from that of a classical particle in free space. This is
due to that Airy packet is not normalizable, and therefore does
not really describe a particle. Nonspreading Airy packets even
exist in time-dependent uniform force \cite{Berry}. It had been
shown that no other potentials except those described above will
support nonspreading Airy packets \cite{Lin}. There are beginning
a lot of investigation on the interesting optical Airy beams. The
Airy optical beams was recently first observed by Christodoulides
\textit{el}. \cite{Christodoulides}. Nonspreading wave packets in
an imaginary potential was also observed by St\"{u}zle
\textit{el}. \cite{Stuzle}. These packets are called Michelangelo
wave packets, which though reach a stationary width, the
probability density however decays in time, and hence are not
really form invariant. In second section, we show that it needs a
real potential to have a form invariant packet.

Nonspreading wave packets is an interesting and important subject.
The essential ingredients of a nonspreading wave packet are its
shape function $f(x)$, its motion $d(t)$, and its phase function
$\phi(x,t)$. From the previous works of many people, we can have a
general discussion on this subject. We derive the general rule for
constructing nonspreading wave packets in the second section. We
derive Berry and Balazs's result in free space and a linear
potential in the third section. We derive Schr\"{o}dinger and
Senitzky's result in the fourth section. We discuss results in other
potentials in the fifth section showing that most stationary
potentials can only support stationary wave packets. We discuss
nonspreading wave packets in time dependent potentials in the final
section.

\section{General derivation}

We start with Schr\"{o}dinger equation,

\BE i\hbar\frac{\partial \Psi(x,t)}{\partial t}= H\Psi(x,t).
\label{b1} \EE\\

\ND The Hamiltonian $H$ is

\BE H =-\frac{\hbar^2}{2m}\partial_x\partial_x+V(x,t). \label{b2}
\EE\\

\ND The potential $V(x,t)$ is allowed to be complex and is written
as

\BE V(x,t)=V_r(x,t)+i\; V_i(x,t),  \label{b3} \EE\\

\ND where $V_r(x,t)$ and $V_i(x,t)$ are two real functions. We are
looking for non-spreading wave packets in a general potential
$V(x,t)$. Let the initial wave packet be

\BE \Psi(x,0)=f(x)\;e^{i\theta(x)},  \label{b4}  \EE\\

\ND where $f(x)$ and $\theta(x)$ are real functions. We require the
time evolved wave function $\Psi(x,t) $ being of the form

\BE \Psi(x,t)=f(q)\;e^{i\phi(x,t)}, \label{b5}  \EE\\

\ND where

 \BE q= x-d(t). \label{b6} \EE\\

\ND The two functions $d(t)$ and $\phi(x,t)$ are real, and satisfy
the boundary conditions

\begin{align}
&d(0)=0.   \nonumber \\
&\phi(x,0)=\theta(x).  \label{b7}  \end{align}\\

\ND The function $d(t)$ describes the motion of the packet, and
$\dot{d}(t)$ is its group velocity. We have $|\Psi(x,0)|^2=f(x)^2$
and $|\Psi(x,t)|^2=f(q)^2$, so $|\Psi(x,t)|^2$ has the same form as
$|\Psi(x,0)|^2$; therefore, $\Psi(x,t)$ is a nonspreading wave
packet.

We need to determine the functional form of $f(q)$ and the related
phase function $\phi(x,t)$. We rewrite the Schr\"{o}dinger equation
as

\BE i\hbar\frac{\partial\Psi(x,t)}{\partial
t}+\frac{\hbar^2}{2m}\partial_x\partial_x\Psi(x,t)-V(x,t)\Psi(x,t)=0.
\label{b8} \EE\\

\ND Substituting (\ref{b5}) into (\ref{b8}) and separating the
equation into real and imaginary parts, we have then two equations.
The real part gives the equation:

\BE \frac{\hbar^2}{2m}f''(q) =[ \;
-V_r(x,t)+\hbar\partial_t\phi(x,t)+
\frac{\hbar^2 (\partial_x\phi(x,t))^2}{2m} \; ] f(q). \label{b9} \EE\\

\ND The imaginary part gives the equation:

\BE [-\hbar \dot{d}(t)+\frac{\hbar^2}{m}\partial_x\phi(x,t)]\;f'(q)
=[\;V_i(x,t)-\frac{\hbar^2}{2m}\partial_x\partial_x\phi(x,t)\;]\;f(q).
\label{b10} \EE\\

\ND We have used the notation: $f'(q)$ represents the derivative of
$f(q)$ with respect to $q$, and $\dot{d}(t)$ the derivative of
$d(t)$ with respect to $t$. The real part equation relates $f''(q)$
and $f(q)$. The imaginary part equation relates $f'(q)$ and $f(q)$.
Eq. (\ref{b10}), when multiplied by $2m f(q),$  can be rearranged
into a form as a flux equation

\BE \partial_x  j(x,t)= 2m V_i(x,t) f(q)^2, \label{b11} \EE\\

where

\BE j(x,t)= [-m \hbar  \; \dot{d}(t)+ \hbar^{2}\;
\partial_x \phi(x,t)]\; f(q)^2 .  \label{b12} \EE\\

\ND Taking integration of $x$ of both sides of (\ref{b11}) from
$-\infty$ to $\infty$, we have

\BE j(\infty,t)-j(-\infty,t)= \int_{-\infty}^{\infty} 2m V_i(x,t)
f(q)^2 dx  \label{b13}. \EE\\

\ND We consider bounded $f(x)$, such that $f(x)\rightarrow 0$ for
$|x|\rightarrow \infty$. Hence the left hand side is zero. The right
hand side, if $V_i$ is not a zero function, in general is not zero.
This means a complex potential in general does not support
nonspreading packets. It is then to require that

\BE
V_i (x,t)=0.\label{b14} \EE\\

\ND From (\ref{b11}), we have

\BE  \partial_x  j(x,t)= 0. \label{b15} \EE\\

\ND (\ref{b15}) shows that $j(x,t)$ can only be a function of time.
Then

\BE j(x,t)= [-m\hbar\dot{d}(t)+\hbar^{2}\;\partial_x
\phi(x,t)]\; f^{2}(q)\equiv b(t). \label{b16} \EE\\

\ND Using again the boundary condition of $f(x)$ at infinity, we
have

\BE
b(t)=0.\label{b17} \EE\\

\ND From (\ref{b16}) and (\ref{b17}), we then have the important
result

\BE -m\hbar\dot{d}(t)+\hbar^{2}\;\partial_x \phi(x,t)=0. \label{b18} \EE\\

\ND From (\ref{b18}), the form of $\phi(x,t)$ is determined to be

\begin{align}
&\phi(x,t)=\phi_1 (t) \; x + \phi_0(t).   \label{b19}\\
&\phi_1 (t)=\frac{m \dot{d}(t)}{\hbar}.  \label{b20}  \end{align}\\

\ND Thus the phase function, $\phi(x,t)$, is restricted to be a
function linear in $x$. From (\ref{b14}), the potential is real, we
then simply write the potential as $V(x,t)$.

We next need to determine the functional form of $f(x)$.
Substituting (\ref{b19}),  (\ref{b20}) into (\ref{b9}), we have the
equation determining the shape of the function $f(q)$.

\BE \frac{\hbar^2}{2m}f''(q) =[\;V(x,t)+
 \frac{m \dot{d}(t)^{2}} {2}
 + m \ddot{d}(t) \;x +\hbar \dot{\phi}_0(t)\;] f(q).
\label{b21}\EE\\

\ND  The left-hand side of (\ref{b21}) is a  function of $q$, so
should be the right-hand side. We replace the variable $x$ by
$q+d(t)$, and then expand the terms in the bracket of the right
hand side in terms of the powers of $q$. We have

\BE \frac{\hbar^2}{2m}f''(q) =(\sum_{n=0}^\infty C_n q^n )\; f(q).\label{b22}\EE\\

\ND Intuitively, each of the coefficient $C_n$ is time dependent
constant, and should be written as $C_n(t)$. However, in order to be
consistent with the left-hand side being a function of $q$ only,
$C_n$ can only be time independent constant. We call the requirement
that $C_n$ should be time independent, the \textbf{consistency}
condition. The constant $C_0$ contains terms $ m \dot{d}(t)^{2}/2
 + m \ddot{d}(t) \;d(t) +\hbar \dot{\phi}_0(t)$, and also the constant term from
$V(x,t)$, which is, $V(d(t),t)$. We denote the coefficient $C_0$ by
$-E_{\textmd{eff}}$ and denote all the rest terms in the summation
by $V_{\textmd{eff}}$. Then

\BE -E_{\textmd{eff}} \equiv V(d(t),t)+\frac{m \dot{d}(t)^{2}} {2}+
m d(t) \ddot{d}(t) + \hbar \dot{\phi}_0(t).
\label{b23}\EE\\

\ND and

\BE V_{\textmd{eff}}(q)\equiv V(q+d(t),t) -V(d(t),t)+ m
\ddot{d}(t)\; q.
\label{b24}\EE \\

\ND Eq. (\ref{b21}) can then be rewritten as a form similar to that
of the time independent Schr\"{o}dinger equation. That is

\BE \;-\frac{\hbar^2}{2m} f''(q) + V_{\textmd{eff}} (q) f(q)\;
= E_{\textmd{eff}} f(q). \label{b25}\EE\\

\ND We call (\ref{b25}) the \textbf{shape} equation. We see that
the function $f(q)$ is the eigenfunction of a Hamiltonian with a
potential $V_{\textmd{\textmd{eff}}}(q)$. Since the functional
form of a nonspreading wave packet does not change in time, the
shape equation being related to the time independent Schrodinger
equation is quite interesting. Finally, by setting
$E_{\textmd{eff}}$ to one of the energy eigenvalues, we can solve
$\phi_0(t)$ from (\ref{b23}).

We have obtained the general rule for constructing nonspreading
wave packets from a general potential $V(x,t)$.  We list the
following formulas for a conclusion:

\begin{align}
&\Psi(x,0)=f(x)\;e^{i\theta(x)}. \label{b26}\\
&\Psi(x,t)=f(q)\;e^{i\phi(x, t)}. \label{b27}\\
&q=x-d(t).\label{b28}\\
&\phi(x,t)=\phi_1 (t) \; x + \phi_0(t). \label{b29} \\
&\phi_1 (t)=\frac{m \dot{d}(t)}{\hbar}. \label{b30}\\
&\;-\frac{\hbar^2}{2m} f''(q) + V_{\textmd{eff}} (q)\;f(q)
=E_{\textmd{eff}} \;f(q). \label{b31}\\
&V_{\textmd{eff}}(q)=\textmd{V}(q+d(t),t) -V(d(t),t)+ m
\ddot{d}(t)\; q. \label{b32}\\
&E_{\textmd{eff}} =-[\;\textmd{V}(d(t),t)+\frac{m \dot{d}(t)^{2}}
{2}+ m \ddot{d}(t) \; d(t)+ \hbar \dot{\phi}_0(t)\;], \label{b33}
\end{align}\\

\ND together with the consistency condition, and the boundary
condition: $f(x)\rightarrow 0$ for $|x|\rightarrow \infty$.

In what follows, we give applications to these formulas.

\section{Berry and Balazs's results}

We start from a real potential $V(x,t)=0$. To determine the shapes
of the nonspreading wave packets in this free space, we first
calculate the effective potential $V_{\textmd{eff}}(q)$. We have
from (\ref{b32})

\BE V_{\textmd{eff}}(q)=m \ddot{d}(t)\; q. \label{c1}\EE\\

\ND Since $ V_{\textmd{eff}}$ can only be a function of $q$, it
needs that $\ddot{d}(t)$ be a constant. We set

\BE \ddot{d}(t)= B^{3}/(2 m^2). \label{c2}\EE\\

\ND Then $V_{\textmd{eff}}= B^{3}q/(2m) $, which is a linear
potential. Substituting this into (\ref{b31}), we obtain the shape
equation

\BE f''(q)- \frac{B^{3}}{\hbar^{2}}\;q\;f(q)= \frac{-2m
E_{\textmd{eff}}}{\hbar^2} f(q). \label{c3}\EE\\

\ND The solution of (\ref{c3}) is the Airy function. Thus the
nonspreading wave packet in free space is determined to be Airy
functions. We consider here the simplest solution by taking
$E_{\textmd{eff}}=0$. The solution of Eq.(\ref{c3}) is then
$A_i[(B/\hbar^{2/3})\; q]$, where we neglect the divergent
solution $B_i[(B/\hbar^{2/3})\;q]$.

The motion of the packet, $d(t)$, can be solved from (\ref{c2}).
Choosing the boundary conditions as $d(0)=0$, and $\dot{d}(0)=0$,
we have the simplest solution:

\BE d(t)= \frac{B^{3}t^{2}}{4 m^{2}}.\label{c4}\EE\\

\ND We obtain $\phi_1(t)$ from (\ref{b30}). Then

\BE \phi_1(t)= \frac{B^{3}t}{2m \hbar}.\label{c5}\EE\\

\ND We obtain  $\phi_0(t)$ from (\ref{b33}) with
$E_{\textmd{eff}}$=0, and the boundary condition $\phi_0 (0)=0$,
we have

\BE \phi_0(t)= -\frac{B^{6} t^{3}}{12 m^{3} \hbar}.\label{c6}\EE\\

\ND We compare these results with Berry's result:

\BE \Psi(x,t)=Ai[\frac{B}{\hbar^{2/3}}(x-\frac{B^3t^2}{4m^2})]
e^{i\frac{B^3t}{2m\hbar}(x-\frac{B^3t^2}{6m^2})}. \label{c7} \EE\\

\ND We see they are the same. From (\ref{c3}), we also see that
the only nonspreading wave packets in free space is the Airy
function \cite{Berry}, \cite{Unnikrishnan}. The trivial plane wave
solution corresponds to taking $B=0$ in (\ref{c3}).

We next consider the real potential

\BE V(x,t)=-F(t)x. \label{d1} \EE\\

\ND From (\ref{b32}), we have

\BE V_{\textmd{eff}}(q)=[-F(t)+ m \ddot{d}(t)] \;q.\label{d2} \EE\\

\ND For consistency, it needs that $[-F(t)+ m \ddot{d}(t)]$ be a
constant. We set

\BE -F(t)+ m \ddot{d}(t)= B^{3}/(2m). \label{d3}\EE\\

\ND Then $V_{\textmd{eff}} = B^{3}/(2m) q $. Thus the nonspreading
wave packets in the potential $ V(x,t)=-F(t)x$ are the Airy
functions again, as found by Berry and Balazs. Again, we calculate

The rest of the complete results are easily shown to be

\BE d(t)= \frac{B^3 t^2}{4m^2}+\frac{1}{m}\int_0^t\int_0^\tau
F(x)dxd\tau. \label{d4}\EE

\BE \phi_1(t)=\frac{B^3 t}{2m \hbar}+
\frac{1}{\hbar}\int_0^tF(\tau)d\tau. \label{d5} \EE

\begin{align}
\phi_0(t)= &-\frac{B^6  t^3}{12m^3 \hbar} -\frac{1}{2 m \hbar}
\int_0^t\; [\int_0^{\tau} F(x) dx\;]^{2}
d\tau \nonumber\\
&-\frac{B^3}{2 m^2 \hbar}\;[\int_0^t \tau \int_0^{\tau} F(x) dx
d\tau +\int_0^t \int_0^{\tau} \int_0^x  F(y) dy dx d\tau].
\label{d6}
\end{align}\\

\ND Using the identity

\BE \int_0^t(t-\tau)F(\tau)d\tau=\int_0^t\int_0^{\tau}F(x)dxd\tau
\label{d7}. \EE\\

\ND we see these results are the same as those of Berry and
Balazs's \cite{Berry}.

\section{Schr\"{o}dinger and Senitzky's results}

We consider the potential

\BE V(x)=\frac{1}{2}m\omega^2x^2.  \label{e1} \EE\\

\ND From (\ref{b32}), we have

\BE V_{\textmd{eff}}(q)=
\frac{1}{2}m \omega ^2 q^2 +m [\;\ddot{d}(t)+ \omega^2 d(t)\;]\; q.\label{e2} \EE\\

\ND It needs that $\ddot{d}+ \omega^2 d(t)$ be a constant. For a
simplest solution, we set the constant to be zero. Hence

\BE \ddot{d}+ \omega^2 d(t)=0.\label{e3} \EE\\

\ND Then $d(t)$ executes simple harmonic motion. This is the same as
the  classical solution of $x(t)$. Taking (\ref{e3}) into
(\ref{e2}), we have

\BE V_{\textmd{eff}}(q)= \frac{1}{2}m \omega ^2
q^2.\label{e4} \EE\\

\ND The shape equation then is of the form

\BE -\frac{\hbar^2}{2m} f''(q) + \frac{1}{2}m \omega ^2 q^2\;f(q)
= E_{\textmd{eff}} f(q)\label{e5}\EE\\

\ND We see $f(q)=\psi_n (q)$, the energy eigenfunctions of the
simple harmonic oscillator. Each eigenfunction $\psi_n (q)$ then
offers a nonspreading wave packet. These packets though with the
shape of stationary energy eigenfunction, however, they are moving
and are in fact executing the simple harmonic motion. We have then
reproduced Schr\"{o}dinger and Senitzky 's result.

Starting from (\ref{e3}), we have the following results

\BE d(t)=x_0 \;sin(\omega t). \label{e6} \EE

\BE \phi_1(t)= \frac{m  \omega x_0}{\hbar}\;cos(\omega t).
\label{e7} \EE

\BE \phi_0(t)=-\frac{m  \omega x_0^2}{4 \hbar} \;sin(2\omega
t) -\frac{E_n t}{\hbar} .\label{e8} \EE\\

We also easily see that if the frequency is time dependent, that is
$\omega= \omega(t)$, then the effective potential from (\ref{e2}) is

\BE V_{\textmd{eff}}(q)=
\frac{1}{2}m \omega (t) ^2 q^2 +m [\;\ddot{d}(t)+ \omega (t)^2 d(t)\;]\; q.\label{e9} \EE\\

\ND From the consistency requirement, we need the coefficients of
$q$ and $q^2$ be constant, hence $\omega$ should be a constant. Thus
a harmonic oscillator with a time dependent frequency can not
support a nonspreading wave packet, as described in Yan's paper
\cite{Yan}.

\section{ potential $ V(x)=\lambda x^n, n\geq 3$}

For potentials of the form

\BE V(x)=\lambda x^n, n \geq 3,  \label{f1} \EE\\

\ND it is trivial to see that from $V_{\textmd{eff}}$ and the
consistency equation, the motion $d(t)$ must be a constant.
function

From (\ref{b32}), we have

\BE V_{\textmd{eff}}(q)=\lambda[q+d(t)]^n -
\lambda d(t)^n +m \ddot{d}\; q. \label{f2} \EE\\

\ND Expanding $\lambda [q+d(t)]^n$ in terms of powers of $q$, we
have the first few terms

\BE \lambda q^n + \lambda n d(t) q^{n-1}+ \cdots. \label{f3} \EE\\

\ND The second term of (\ref{f3}) shows that $d(t)$ needs to be a
constant. Thus the packet does not move. This implies that the
nonspreading packets can only be stationary packets. We set this
constant as $d$, then we have

\BE d(t)=d. \label{f4} \EE\\

\ND And then we have

\BE V_{\textmd{eff}}(q)=\lambda[q+d)]^n -
\lambda d^n . \label{f5} \EE\\

\ND Substituting this to (\ref{b31}), we have the shape equation of
the form

\BE -\frac{\hbar^2}{2m} f''(q) + \lambda [q+d)]^n \;f(q) = E f(q),
\label{f6} \EE\\

\ND where $E=E_{eff}+\lambda d^n$. As $q+d=x$, formula (\ref{f6})
shows that the function $f(q) = \psi_n(x)$, the energy eigenfunction
of the Hamiltonian with potential $V(x)=\lambda x^n$. Since the
packet is the energy eigenfunction, it is a stationary wave; and
therefore it does not move. We conclude that there is no moving
nonspreading packet in a potential of the form $V(x,t)=\lambda x^n,
n \geq 3$. The same argument applies also to showing that a
potential, such as $V(x)=\lambda/x$, does not support moving
nonspreading wave packets. We see most stationary potentials can
only support stationary wave packets.

\section{ time dependent potentials with moving nonspreading wave packets}

From above results, hence, in order to have moving nonspreading wave
packets, we need to consider time dependent potentials. The simplest
way is to move the potential. That is to consider $V(x-d(t))$. In
this way, there are in fact many nonspreading wave packets with
arbitrary motion that can be constructed. For instance, we start
from a given motion, such as $d(t)= t^2$, etc., and then consider
the time dependent potential

\BE V(x,t)=\frac{1}{2} m \omega^2 (x-d(t))^2 - m x \ddot{d}(t)
-\frac{1}{2} m \omega^2 x_0 ^2 \;sin(\omega t)^2.
\label{g1} \EE\\

\ND The effective potential calculated from (\ref{g1}) and
(\ref{b32}) is

\BE V_{\textmd{eff}}(q)= \frac{1}{2} m \omega^2 q^2. \label{g2} \EE\\

\ND Thus, the potential in Eq. (\ref{g1}) offers nonspreading wave
packets, with the shapes of the eigenfunctions of SHO, moving with
the arbitrary function $d(t)$. If we set $d(t)= x_0 \; sin(\omega
t)$, and substituting this to (\ref{g1}), we find the needed
potential for this motion is $V(x,t)=\frac{1}{2} m \omega^2 x^2$, as
expected.

Another example is considering the time dependent potential

\BE V(x,t)= \lambda (x-d(t))^4 - m x \ddot{d}(t).
\label{g3} \EE\\

\ND The effective potential can be calculated to be

\BE V_{\textmd{eff}}(q)= \lambda \; q^4 . \label{g4} \EE\\

\ND The potential in Eq. (\ref{g3}) then offers nonspreading wave
packets with the shapes of the eigenfunctions of the potential
$V(x)=\lambda x^4$. These nonspreading wave packets move with the
arbitrary function $d(t)$ as the moving potential does. We see that
in order to move a nonspreading wave packet, we need a moving
potential, and also a linear potential $- m x \ddot{d}(t)$, which
offers a force $ m \ddot{d}(t)$; this is just the force needed for a
particle with a motion $d(t)$. In conclusion, a moving potential
$\lambda (x-d(t))^4$, together with a time dependent linear
potential $- m x \ddot{d}(t)$, will drive stationary-shaped packets
going around with a journey $d(t)$. It is like to have water moving
in space, we put it in a bowl, and carry the bowl going around.

\newpage

\Large

\end{document}